\documentclass[iop]{emulateapj-rtx4}

\usepackage{mathrsfs }
\usepackage{natbib}
\usepackage{amssymb}
\usepackage{ulem} 
\newcommand{\secpoint}{\mbox{$''\mskip-7.6mu.\,$}}

\newcommand {\kms}    {km~s$^{-1}$}

\begin{document}

\title{The MOSDEF Survey: First Measurement of Nebular Oxygen Abundance at $\lowercase{z}>4$\altaffilmark{1}}

\author{
 Alice E. Shapley,\altaffilmark{2}
 Ryan L. Sanders,\altaffilmark{2}
 Naveen A. Reddy,\altaffilmark{3}	
 Mariska Kriek,\altaffilmark{4}	
 William R. Freeman,\altaffilmark{3}	
 Bahram Mobasher,\altaffilmark{3}	
 Brian Siana,\altaffilmark{3}
 Alison L. Coil,\altaffilmark{5}
 Gene C. K. Leung,\altaffilmark{5}
 Laura deGroot,\altaffilmark{6}
 Irene Shivaei,\altaffilmark{3}
 Sedona H. Price,\altaffilmark{4}
 Mojegan Azadi,\altaffilmark{5}
 James Aird\altaffilmark{7}
 }

\altaffiltext{1}{Based on data obtained at the W.M. Keck Observatory, which is operated as a scientific partnership among the California Institute of Technology, the University of California,  and the National Aeronautics and Space Administration, and was made possible by the generous financial support of the W.M. Keck Foundation.}
\altaffiltext{2}{Department of Physics and Astronomy, University of California, Los Angeles, 430 Portola Plaza, Los Angeles, CA 90095, USA}
\altaffiltext{3}{Department of Physics and Astronomy, University of California, Riverside, 900 University Avenue, Riverside, CA 92521, USA}
\altaffiltext{4}{Astronomy Department, University of California at Berkeley, Berkeley, CA 94720, USA}
\altaffiltext{5}{Center for Astrophysics and Space Sciences, Department of Physics, University of California, San Diego, 9500 Gilman Drive., La Jolla, CA 92093, USA}
\altaffiltext{6}{Department of Physics, The College of Wooster, 1189 Beall Avenue, Wooster, OH 44691, USA}
\altaffiltext{7}{Institute of Astronomy, University of Cambridge, Madingley Road, Cambridge CB3 0HA, UK}
\email{aes@astro.ucla.edu}

\shortauthors{Shapley et al.}


\shorttitle{Nebular Oxygen Abundance at $z>4$}


\begin{abstract} 
We present the first spectroscopic measurement of multiple rest-frame optical emission lines
at $z>4$. During the MOSFIRE Deep Evolution Field (MOSDEF) survey, we observed
the galaxy GOODSN-17940 with the Keck~I/MOSFIRE spectrograph. The $K$-band spectrum of GOODSN-17940 
includes significant detections of the [O{\sc ii}]$\lambda\lambda 3726,3729$, [Ne{\sc iii}]$\lambda3869$,
and H$\gamma$ emission lines and a tentative detection of H$\delta$, indicating
$z_{\rm{spec}}=4.4121$. GOODSN-17940 is an actively star-forming $z>4$ galaxy
based on its $K$-band spectrum and broadband spectral energy distribution.
A significant excess relative to the surrounding continuum
is present in the {\it Spitzer}/IRAC channel~1 photometry of GOODSN-17940, due primarily
to strong H$\alpha$ emission with a rest-frame equivalent width of $\mbox{EW(H}\alpha)=1200$~\AA. 
Based on the assumption of $0.5 Z_{\odot}$ models and the Calzetti
attenuation curve, GOODSN-17940 is characterized by $M_*=5.0^{+4.3}_{-0.2}\times 10^9 M_{\odot}$. The Balmer
decrement inferred from H$\alpha$/H$\gamma$ is used to dust correct the H$\alpha$ emission, yielding
$\mbox{SFR(H}\alpha)=320^{+190}_{-140} M_{\odot}\mbox{ yr}^{-1}$. These $M_*$ and SFR values place GOODSN-17940
an order of magnitude in SFR above the $z\sim 4$ star-forming ``main sequence." 
Finally, we use the observed ratio of [Ne{\sc iii}]/[O{\sc ii}] to estimate the nebular oxygen abundance in GOODSN-17940,
finding $\mbox{O/H}\sim 0.2 \mbox{ (O/H)}_{\odot}$. Combining our new [Ne{\sc iii}]/[O{\sc ii}] measurement with 
those from stacked spectra at $z\sim 0, 2, \mbox{ and } 3$, we show that GOODSN-17940 represents
an extension to $z>4$ of the evolution towards higher [Ne{\sc iii}]/[O{\sc ii}] (i.e., lower $\mbox{O/H}$)
at fixed stellar mass. It will be possible to perform the measurements
presented here out to $z\sim 10$ using the {\it James Webb Space Telescope}.

\end{abstract} 



\section{Introduction}
\label{sec:introduction}

The study of high-redshift galaxies has been transformed by major
recent advances in infrared instrumentation. New multi-object
near-IR spectrographs on 8-10-meter class ground-based telescopes have
enabled ground-breaking discoveries about galaxies at $z\sim 1.5-3.5$,
the epoch of peak star-formation and black hole accretion activity \citep{madau2014}.
In particular, statistical samples of rest-frame
optical spectra collected with the MOSFIRE spectrograph on Keck~I \citep{mclean2012}
and KMOS spectrograph  on the VLT \citep{sharples2013}
have yielded detailed probes of the gas, dust, heavy
elements, stars, supermassive black holes, and dynamics in galaxies at
high redshift -- all of which are essential to understanding the basic
questions of galaxy formation.

However, there are virtually {\it no} direct rest-frame optical spectroscopic
measurements beyond $z=4$. At such redshifts, the strongest rest-frame optical
emission lines (e.g., H$\alpha$ and [O{\sc iii}]$\lambda5007$) are shifted into the thermal infrared 
and inaccessible from the ground. The strengths of these lines have only been inferred indirectly
at $z\sim 4-7$ based on {\it Spitzer}/IRAC photometric measurements \citep[e.g.,][]{stark2013,smit2014}.
In addition, spectroscopic detections of the [O{\sc ii}]$\lambda\lambda 3726,3729$ doublet have been
presented for only three gravitationally lensed galaxies, all at $z=4.9$ \citep{troncoso2014,swinbank2007,swinbank2009}.

The launch of the {\it James Webb Space Telescope}
({\it JWST}) will enable the extension of rest-frame optical spectroscopic probes of the contents
of galaxies from the peak epoch of star formation out to $z=4-10$. Meanwhile,
in the course of the MOSFIRE Deep Evolution Field (MOSDEF) survey \citep{kriek2015}, 
we have serendipitously collected the rest-frame optical spectrum of an unlensed
galaxy in the GOODS-N field at $z=4.4121$.
The analysis of this galaxy, GOODSN-17940,
provides the {\it first} rest-frame optical nebular abundance estimate at $z>4$ and
a preview of the advances soon to come with {\it JWST}.
In \S\ref{sec:observations}, we describe our observations and data analysis.
In \S\ref{sec:spectrum}, we present the MOSFIRE $K$-band spectrum of GOODSN-17940,
and determine several key physical properties including the nebular oxygen abundance
in \S\ref{sec:physprop}.  Finally, in \S\ref{sec:discussion}, we consider an AGN contribution to GOODSN-17940,
place its oxygen abundance measurement into the context of the global chemical evolution of star-forming galaxies, and consider
future rest-frame optical spectroscopic observations of GOODSN-17940 and other $z>4$ star-forming galaxies.
Throughout, we adopt cosmological parameters of
$H_0=70 \mbox{ km  s}^{-1} \mbox{ Mpc}^{-1}$, $\Omega_M = 0.30$, and
$\Omega_{\Lambda}=0.7$.

\begin{figure*}[t!]
\centering
\includegraphics[width=0.95\textwidth]{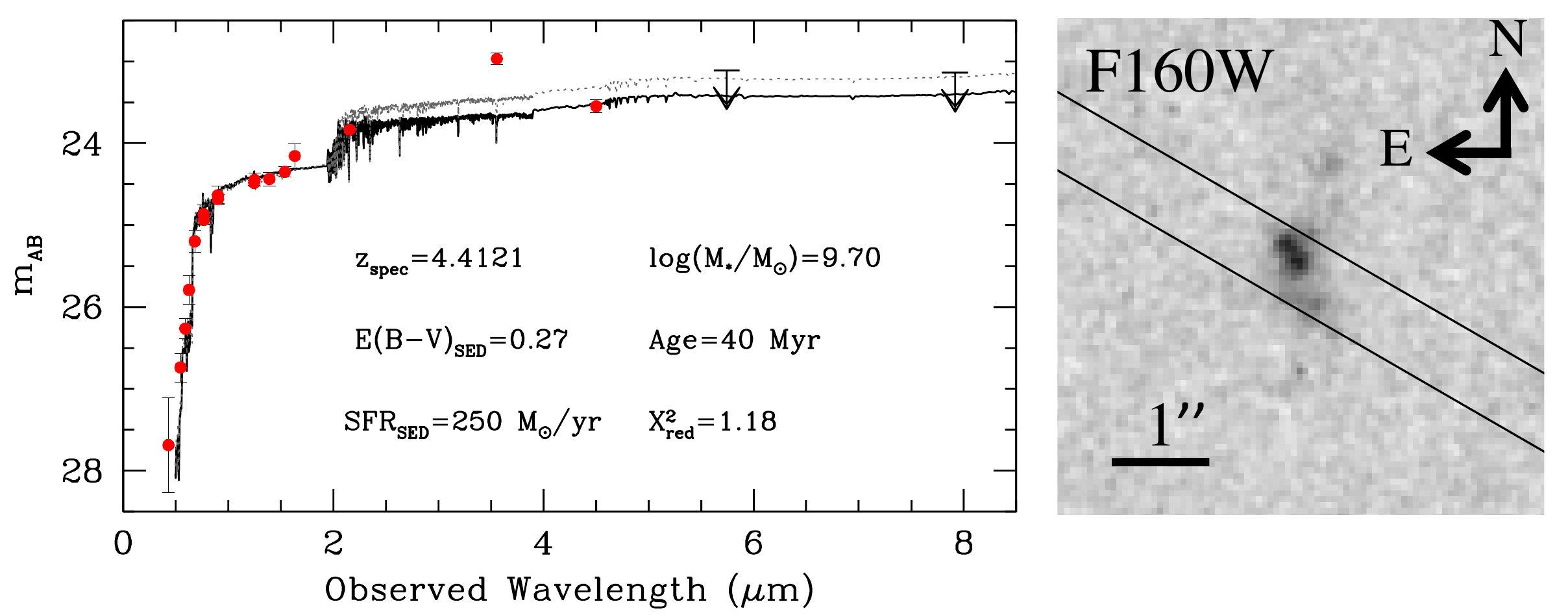}
\caption{Stellar populations and image of GOODSN-17940.
{\bf Left:} The observed and best-fit model SEDs for GOODSN-17940.
Photometric detections are indicated with red symbols, including the IRAC channel~1 datapoint,
which suggests significant contamination by nebular emission.
We plot 2$\sigma$ limits for the IRAC channel~3 and 4 non-detections.
The best-fit stellar population model of GOODSN-17940, excluding IRAC channel~1, is shown
as the solid black curve. We also list the parameters of this model from \citet{conroy2009}, which assumes $Z=0.5 Z_{\odot}$
and a Calzetti attenuation law.
The best-fit model to the SED including IRAC channel~1 is shown as a dotted grey curve.
This fit is biased high by a factor of $2.7$ in stellar mass relative to the clean fit.
{\bf Right:} WFC3/F160W postage stamp image of GOODSN-17940, with the 0\secpoint7 MOSFIRE slit overlaid.
At $z=4.4121$, the F160W filter probes $\lambda_{\rm{rest}}=2840$~\AA.
GOODSN-17940 consists of two central clumps separated by $\sim 1$~kpc
(proper), surrounded by an irregular, lower-surface brightness region that extends towards the south.
The emission located roughly 1'' to the northwest is identified as a separate
object in the 3D-HST photometric catalog (ID=17884, $z_{\rm{phot}}=3.93$).
}
\label{fig:17940-sedhst}
\end{figure*}

\section{Observations}
\label{sec:observations}

GOODSN-17940 (R.A.=12:36:35.49 and decl.=62:13:50.0 (J2000), $m_{\rm{F160W,AB}}=24.35$)
was observed as part of the MOSDEF survey \citep{kriek2015}. MOSDEF is
a large survey of the rest-frame optical spectra of $\sim 1500$ galaxies
at $1.4 \leq z \leq 3.8$. We collected Keck~I/MOSFIRE data
for MOSDEF over 48.5 nights spanning from 2012 December
through 2016 May. MOSFIRE $H$- and $K$-band spectroscopic data were obtained for 
a multi-object slitmask containing GOODSN-17940 on 2016 March 18. 
The slitwidth was 0\secpoint7, yielding a spectral resolution of $\sim 3650$
in $H$ and $\sim 3600$ in $K$.  The total integration time in both $H$ and $K$ was 120 minutes.
The conditions were clear with a median seeing of 0\secpoint6.

We reduced the raw data to produce two-dimensional science and error spectra
using a custom IDL pipeline. 
We then optimally extracted one-dimensional science and error spectra from the two-dimensional spectra,
applying slit-loss corrections as described in \citet{kriek2015}. 
Our interpretation of GOODSN-17940 is greatly enhanced by the extensive
multi-wavelength coverage in the GOODS-N field, including broadband photometry
in 22 filters spanning from the near-UV through mid-infrared 
\citep{skelton2014,momcheva2016}.

MOSDEF targets galaxies in three
discrete redshift intervals, $1.37\leq z\leq 1.70$, $2.09 \leq z \leq 2.61$,
and $2.95\leq z \leq 3.80$ \citep{kriek2015}. GOODSN-17940 was targeted based on a prior
spectroscopic redshift of $z_{\rm{spec}}=3.665$, which placed it in the most distant
MOSDEF target redshift interval. This redshift was derived from
Keck/DEIMOS spectroscopy and the putative detection of Ly$\alpha$ at 5673~\AA,
which was unconfirmed when recently re-examined
(S. Hemmati, private communication). The photometric redshift of GOODSN-17940 in
the publicly-available 3D-HST catalog is $z_{\rm{phot}}=4.447$ \citep{momcheva2016},
consistent with the one measured from our MOSFIRE $K$-band spectrum. 
We also note that another  Keck/DEIMOS spectrum nominally targeting
the fainter, diffuse morphological component 0\secpoint6 to the south (see Figure~\ref{fig:17940-sedhst}, right)
yielded a secure detection of Ly$\alpha$ at $z = 4.416$ (M. Dickinson, private communication).

\section{The Spectrum of GOODSN-17940}
\label{sec:spectrum}

\begin{figure*}
\centering
\includegraphics[width=0.65\textwidth]{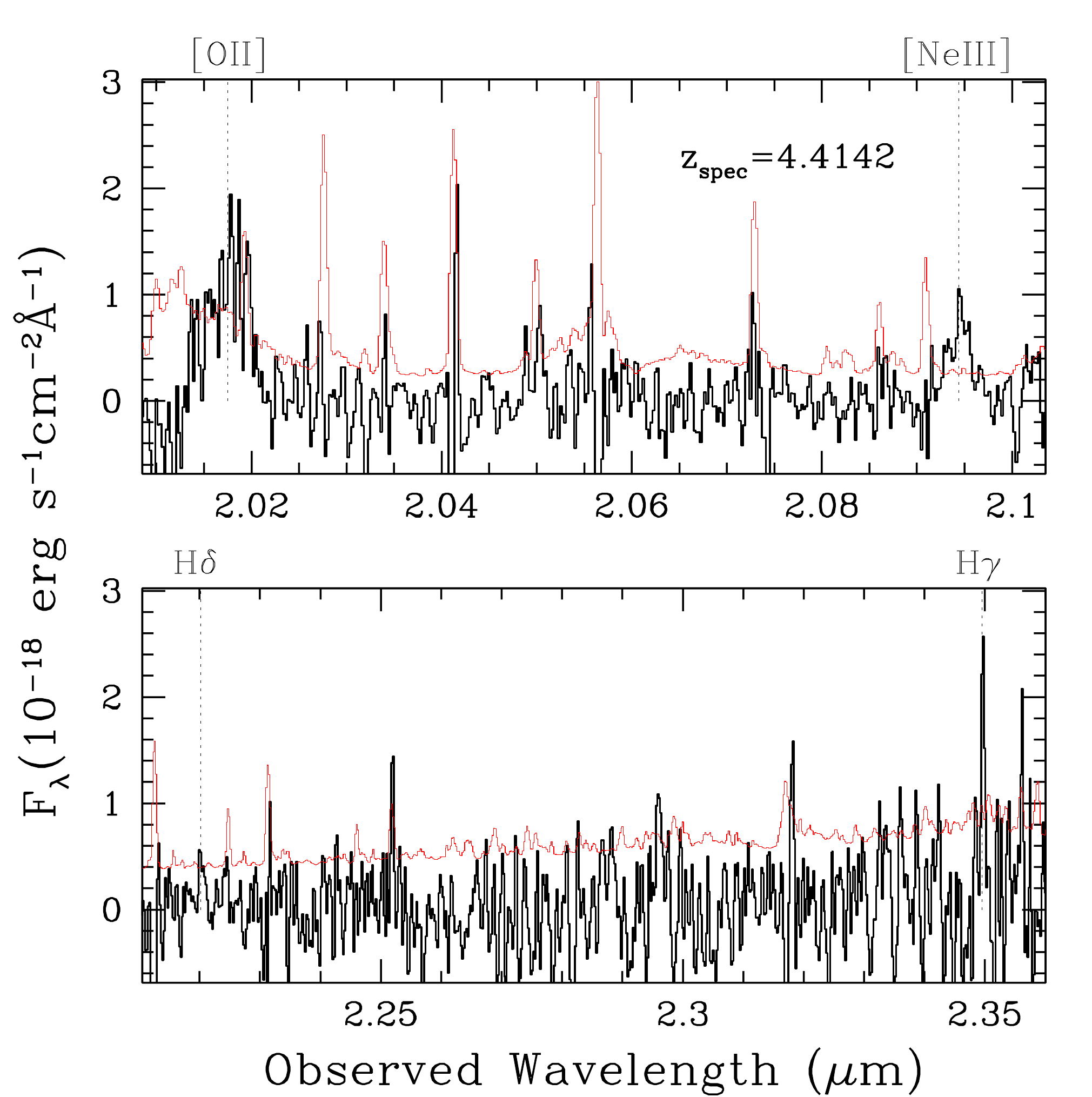}
\caption{Observed-frame MOSFIRE $K$-band spectrum of GOODSN-17940.
Flux density units are $10^{-18} \mbox{ ergs s}^{-1}\mbox{ cm}^{-2}\mbox{\AA}^{-1}$.
The error spectrum is overplotted in red, and the locations
of [O{\sc ii}]$\lambda\lambda 3726,3729$, [Ne{\sc iii}]$\lambda 3869$,
H$\delta$, and H$\gamma$ are labeled and marked
with dotted vertical lines. The redshift we report
from the highest S/N feature, [Ne{\sc iii}],
is consistent with those estimated from the other
labeled features.
}
\label{fig:17940spec-mosfire}
\end{figure*}

As shown in Figure~\ref{fig:17940spec-mosfire}, the MOSFIRE $K$-band
spectrum of GOODSN-17940 includes significant detections of the [O{\sc ii}]$\lambda\lambda3726,3729$,
[Ne{\sc iii}]$\lambda 3869$, and H$\gamma$ emission lines, along with a tentative $2.3\sigma$
detection of H$\delta$. Based on the observed centroid of the highest signal-to-noise (S/N)
feature, [Ne{\sc iii}]$\lambda 3869$, we estimate a spectroscopic redshift of $z_{\rm{spec}}=4.4121$.
We fit Gaussian profiles to the emission lines, using a single-component model for [Ne{\sc iii}]$\lambda 3869$
and the Balmer lines, and a two-component model for [O{\sc ii}]$\lambda\lambda3726,3729$. The ratio
of wavelengths was fixed for the two [O{\sc ii}] doublet members, and their profile FWHMs were forced
to be identical. Uncertainties on reported quantities were estimated using Monte Carlo
simulations, in which 500 perturbed fake spectra were generated from the observed spectrum
according to the error spectrum, and the measurements repeated on each fake spectrum.
The $1\sigma$ confidence intervals for each measurement were determined from the distributions
of simulated best-fit parameters.

Observed line fluxes and uncertainties are listed in Table~\ref{tab:emline}.
We note that both the [Ne{\sc iii}] and [O{\sc ii}] emission lines are well-resolved spectrally,
with the [Ne{\sc iii}] line suggesting a line width (corrected for instrumental broadening) of $\sigma_v= 160^{+10}_{-50}$\kms.
The Balmer features are not well-resolved in the spectral dimension, which may be a result of
their lower S/N. Finally, while the density-sensitive [O{\sc ii}] doublet is blended and only of $4.7\sigma$ significance,
our two-component fitting simulations suggest that the bluer doublet member is the stronger one, corresponding
to the high-density limit \citep{sanders2016a}.

\begin{deluxetable}{lc}
\tabletypesize{\footnotesize}
\tablecolumns{3}
\tablewidth{2in}
\tablecaption{Emission-line Properties of GOODSN-17940\tablenotemark{a}\label{tab:emline}}
\tablehead{
\colhead{Line} & \colhead{Flux} \\
}
\startdata
$\mbox{[O{\sc ii}]}\lambda\lambda3726,3729$ & $6.9\pm1.5$ \\
$\mbox{[Ne{\sc iii}]}\lambda3869$           & $2.2\pm 0.5$ \\
H$\delta$                      & $0.7\pm0.3$ \\
H$\gamma$                      & $2.1 \pm 0.5$ 
\enddata
\tablenotetext{a}{Observed emission line flux in units of $10^{-17}\mbox{ ergs s}^{-1}\mbox{ cm}^{-2}$.}
\end{deluxetable}
\vspace{0.1cm}

\section{The Physical Properties of GOODSN-17940}
\label{sec:physprop}

The combination of the MOSFIRE $K$-band spectrum, broadband spectral
energy distribution (SED), and {\it HST} imaging of GOODSN-17940 provides a window into a remarkable $z\sim 4$
galaxy, including its stellar population, dust content, current star-formation rate (SFR),
and nebular oxygen abundance.

\subsection{Stellar Population}
\label{sec:physprop-sedfit}
Figure~\ref{fig:17940-sedhst} (left) shows the observed
SED of GOODSN-17940. There is evidence for
strong H$\alpha$+[N{\sc ii}]$\lambda\lambda 6548,6584$+[S{\sc ii}]$\lambda\lambda6717,6731$ 
emission-line contamination in the {\it Spitzer}/IRAC channel~1 datapoint at 3.6~$\mu$m, as 
this measurement is significantly elevated relative to the surrounding datapoints
at $K$ band and IRAC channel~2. In order to quantify both stellar population parameters
and nebular emission flux within IRAC channel~1, we modeled the broadband SED of GOODSN-17940 using the fitting
code, FAST \citep{kriek2009}. In this analysis, we excluded the contaminated IRAC channel~1
datapoint,  and adopted the population synthesis models of \citet{conroy2009},
the \citet{calzetti2000} attenuation curve, and a \citet{chabrier2003} IMF. We further assumed
a delayed exponential star-formation history of the form $\mbox{SFR}\propto t \exp(-t/\tau)$, 
with $t$ the time since the onset of star formation and $\tau$ the characteristic
timescale for the decay of star formation. Based on the inferred oxygen abundance of GOODSN-17940 described 
in Section~\ref{sec:physprop-oh}, we used model spectra with 0.5 solar
abundance ($Z=0.5Z_{\odot}$), the lowest-metallicity default option in FAST for the \citet{conroy2009} models.

The SED modeling for GOODSN-17940 results in best-fit parameters of $M_*=5.0^{+4.3}_{-0.2}\times10^9 M_{\odot}$,
$E(B-V)_{\rm{SED}}=0.27^{+0.0}_{-0.05}$, SFR$_{\rm{SED}}$=$250^{+20}_{-140}\mbox{ }M_{\odot}\mbox{ yr}^{-1}$,
and a young age ($\leq 200$~Myr). Figure~\ref{fig:17940-sedhst} shows the best-fit
models both excluding (black, solid) and including (grey, dashed) IRAC channel~1. In
the latter case, the best-fit stellar mass is biased towards a higher value of $1.3\times 10^{10}M_{\odot}$.
We note that the best-fit stellar population parameters for GOODSN-17940 depend systematically
on the input model assumptions. For example, if we instead follow \citet{reddy2017} and fit GOODSN-17940 with a
$Z=0.14 Z_{\odot}$ BPASS model \citep{stanway2016} that includes massive stellar binaries,
nebular continuum emission, and a constant star-formation history, and we assume
an SMC attenuation curve \citep{gordon2003}, we find $M_*=1.3\times10^{10} M_{\odot}$,
$E(B-V)_{\rm{SED}}=0.10$, SFR$_{\rm{SED}}$=$30\mbox{ }M_{\odot}\mbox{ yr}^{-1}$,
and Age=$500\mbox{ Myr}$. Accordingly, these other assumptions yield
a higher stellar mass for GOODSN-17940, and lower
SFR and specific SFR ($SFR/M_*$). More work is needed
to determine the optimal SED modeling assumptions for $z>4$ galaxies such as GOODSN-17940.
In what follows, we interpret GOODSN-17940 using the $Z=0.5Z_{\odot}$ \citeauthor{conroy2009}
model, but highlight systematic uncertainties related to SED fitting when relevant.

\subsection{H$\alpha$ Emission}
\label{sec:physprop-ha}
The observed excess at IRAC channel~1 relative to the best-fit SED model
excluding this datapoint can be used to estimate the nebular emission
flux, $F_{\rm{em}}$, causing the excess. Specifically:

\begin{equation}
F_{\rm{em}}=\Delta\lambda(F_{\lambda,\rm{ch1,obs}}-F_{\lambda,\rm{ch1,model}})
\label{eq:fem}
\end{equation}

where $\Delta\lambda=7500$~\AA\ is the width of the IRAC channel~1 filter,
$F_{\lambda,\rm{ch1,obs}}=5.53^{+0.03}_{-0.03}\times 10^{-20}\mbox{ ergs s}^{-1}\mbox{ cm}^{-2}\mbox{\AA}^{-1}$ is the observed IRAC channel~1 flux density
and $F_{\lambda,\rm{ch1,model}}=2.93^{+0.03}_{-0.01}\times 10^{-20}\mbox{ ergs s}^{-1}\mbox{ cm}^{-2}\mbox{\AA}^{-1}$
is the IRAC channel~1 flux density in the best-fit SED model excluding
the IRAC channel~1 datapoint. Based on the oxygen abundance in the ionized ISM of GOODSN-17940 discussed below, we further
assume that the nebular emission is dominated by H$\alpha$, and make the approximation
$F_{\rm{em}} \approx F(\mbox{H}\alpha)$. We find that $F(\mbox{H}\alpha)=2.0^{+0.2}_{-0.3} \times 10^{-16}\mbox{ ergs s}^{-1}\mbox{ cm}^{-2}$,
which corresponds to a rest-frame equivalent width of $EW(\mbox{H}\alpha)=1200^{+100}_{-200}$~\AA. This EW(H$\alpha$)
is significantly larger than the average of $\sim 400$~\AA\ found by \citet{stark2013} for a sample
of 92 spectroscopically-confirmed star-forming galaxies at $3.8 \leq z \leq 5.0$, and demonstrates
the extreme nature of GOODSN-17940. Based on our analysis,
the contribution from nebular line emission in GOODSN-17940 is slightly less than $50$\% of the total flux 
in the IRAC channel~1 band. Both the inferred $F(\mbox{H}\alpha)$ and $EW(\mbox{H}\alpha)$ are insensitive
to the choice of SED fitting assumptions.

\subsection{Dust Extinction}
\label{sec:physprop-dust}
We combine $F(\mbox{H}\alpha)$ inferred from the IRAC channel~1 excess,
and $F(\mbox{H}\gamma)$ measured directly from the MOSFIRE $K$-band spectrum,
to estimate the Balmer decrement.
The theoretical $F(\mbox{H}\alpha)$/$F(\mbox{H}\gamma)$ ratio is 6.11,
based on the assumption of a temperature of $T=10,000$~K, an electron density
of $n_{\rm{e}} = 100$~cm$^{-3}$, and Case B recombination.
The observed ratio of $F(\mbox{H}\alpha)$/$F(\mbox{H}\gamma)_{\rm{obs}}=9.3^{+3.1}_{-2.4}$
then implies $E(B-V)_{\rm{gas}}=0.28\pm 0.19$, with the assumption of the \citet{cardelli1989}
dust attenuation curve. In this calculation, we have not corrected H$\alpha$
and H$\gamma$ fluxes for underlying stellar Balmer absorption. Based on the best-fit
stellar population model for GOODSN-17940, we find that the H$\gamma$ absorption correction is
$\sim 1$\% of the measured H$\gamma$ emission flux, while the fractional contribution
of the Balmer absorption correction to H$\alpha$ is an order of magnitude smaller. Therefore, the
application of such corrections makes no significant difference to our results.

\begin{figure*}[t!]
\centering
\includegraphics[width=0.95\textwidth]{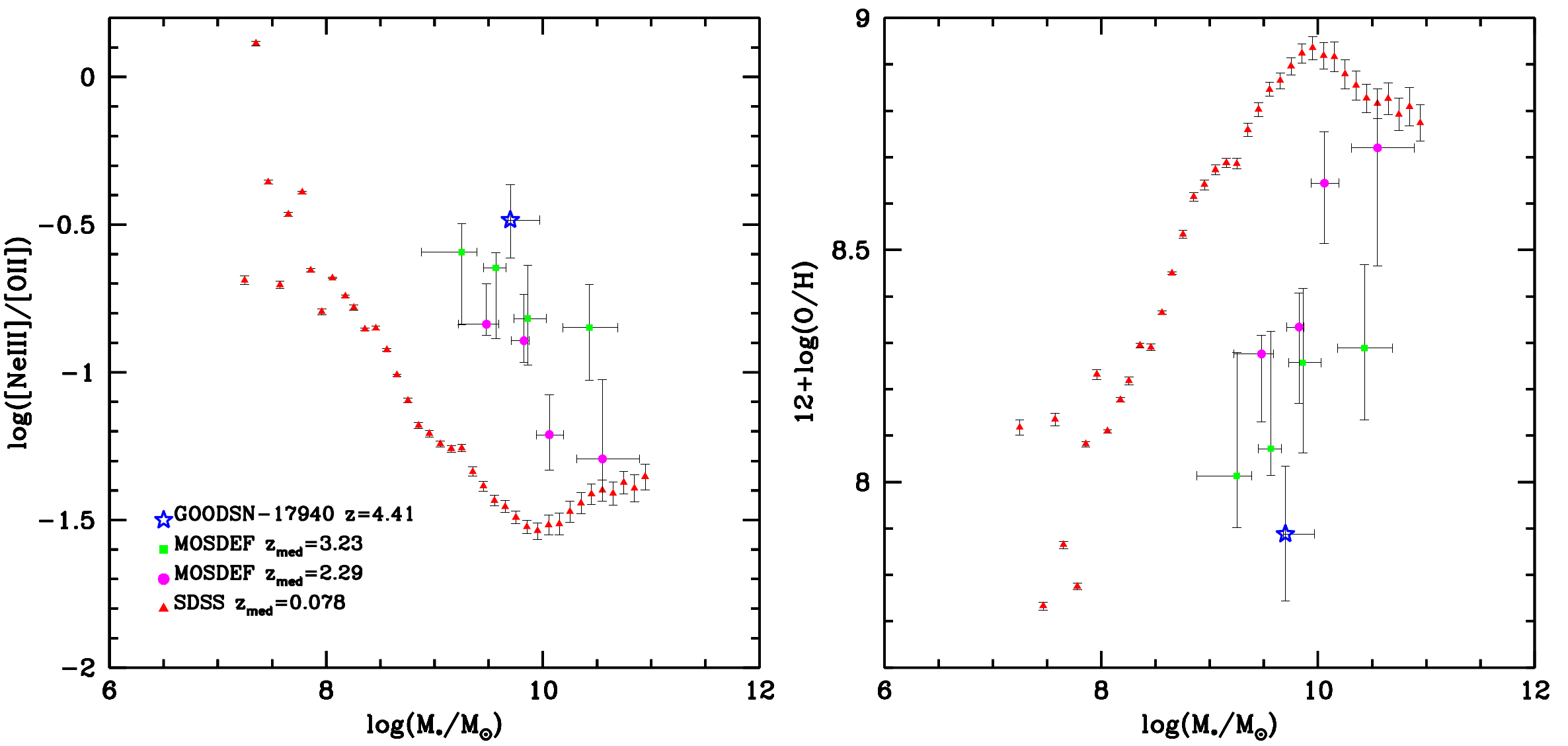}
\caption{Evolution in the Mass-Metallicity Relationship. {\bf Left:} 
The observed ratio of [Ne{\sc iii}]/[O{\sc ii}] vs. stellar mass for composite spectra
of SDSS galaxies of stellar mass at $z\sim 0$ (red triangles);
composite spectra of MOSDEF samples at $z\sim 2$ and $z\sim 3$ (magenta circles and green
squares, respectively); and GOODSN-17940
at $z=4.4121$ (large blue star). There is a clear progression towards higher [Ne{\sc iii}]/[O{\sc ii}]
at fixed mass, which can be understood as an evolution towards lower oxygen abundance.
{\bf Right:} $12+\log(\mbox{O/H})$ vs. stellar mass, based on the [Ne{\sc iii}]/[O{\sc ii}] calibration of \citet{maiolino2008}.
As a single, high-specific-SFR galaxy, GOODSN-17940 may not be representative
of the average $5\times 10^9 M_{\odot}$ galaxy at $z=4.4$, and its offset relative
to the lower-redshift systems of similar mass is likely enhanced relative to more typical
$z>4$ galaxies.
}
\label{fig:ne3o2-m-panels}
\end{figure*}

\subsection{Current Star Formation}
\label{sec:physprop-sfr}
We apply the inferred $E(B-V)_{\rm{gas}}$ to calculate the dust-corrected H$\alpha$ flux
and SFR. The dust-corrected H$\alpha$ flux is
$F(\mbox{H}\alpha)=3.7^{+2.0}_{-1.6} \times 10^{-16}\mbox{ ergs s}^{-1}\mbox{ cm}^{-2}$,
which corresponds to $SFR(\mbox{H}\alpha)=320^{+190}_{-140}\mbox{ }M_{\odot}\mbox{ yr}^{-1}$
at $z=4.4121$. Here we assume the \citet{kennicutt1998a} calibration between H$\alpha$ luminosity
and SFR, applying a normalization factor of 1.8 to convert to a \citet{chabrier2003}
IMF. The \citeauthor{kennicutt1998a} calibration likely overestimates $SFR(\mbox{H}\alpha)$ by up to a factor of $\sim 2$ for
GOODSN-17940, given that this calibration is based on solar metallicity models, while the metallicity we infer in
Section~\ref{sec:physprop-oh} for GOODSN-17940 is sub-solar. As shown in \citet{stanway2016}, sub-solar
stellar population models produce more ionizing photons per unit solar mass of star formation.
Regardless, SFR(H$\alpha$) is consistent within $1\sigma$ with
SFR$_{\rm{SED}}=250^{+20}_{-140}\mbox{ }M_{\odot}\mbox{ yr}^{-1}$
inferred from the best-fit stellar population model. Both methods indicate
that GOODSN-17940 is highly star-forming, and significantly elevated in SFR relative to the
``main sequence" correlation between SFR and $M_*$ at $z=4-5$ \citep[e.g.,][]{salmon2015}.
GOODSN-17940 instead falls within the $z=4-5$ ``starburst" cloud recently identified by \citet{caputi2017}.

\subsection{Nebular Oxygen Abundance}
\label{sec:physprop-oh}
As highlighted by \citet{nagao2006}, the [Ne{\sc iii}]$\lambda 3869$/[O{\sc ii}]$\lambda\lambda 3726,3729$ ratio declines
monotonically with increasing metallicity.
The measurement of both [Ne{\sc iii}]$\lambda 3869$ and [O{\sc ii}]$\lambda\lambda 3726,3729$
can therefore be used to estimate the nebular oxygen abundance in GOODSN-17940.
Both \citet{maiolino2008} and \citet{jones2015} have presented calibrations between 
[Ne{\sc iii}]/[O{\sc ii}] and $12+\log(\mbox{O/H})$. \citet{jones2015} restricts the calibration of [Ne{\sc iii}]/[O{\sc ii}] 
to the low-metallicity regime in which direct oxygen abundances are available ($12+\log(\mbox{O/H})\leq 8.4$),
while \citet{maiolino2008} extends this calibration into the metal-rich regime using
$12+\log(\mbox{O/H})$ estimates based on photoionization modeling. The measured [Ne{\sc iii}]/[O{\sc ii}]
ratio is $F(\mbox{[Ne{\sc iii}]})/F(\mbox{[O{\sc ii}]})=0.33^{+0.10}_{-0.08}$, not including a dust
correction due to the proximity of [Ne{\sc iii}]$\lambda 3869$ and [O{\sc ii}]$\lambda\lambda 3726,3729$ in
wavelength. This ratio corresponds to
$12+\log(\mbox{O/H})= 7.89^{+0.15}_{-0.14} \; [8.05^{+0.06}_{-0.06}]$ according to the calibrations of
\citeauthor{maiolino2008} [\citeauthor{jones2015}]. 
Both [Ne{\sc iii}]/[O{\sc ii}] calibrations suggest
that GOODSN-17940 has $\mbox{O/H}\sim 0.2 \mbox{ (O/H)}_{\odot}$ \citep{asplund2009}.

There has been much recent discussion in the literature regarding the applicability
(or lack thereof) of strong-line metallicity calibrations at $z\geq 2$, in particular
those including the [N{\sc ii}]$\lambda 6584$/H$\alpha$ ratio \citep{steidel2014,shapley2015,
sanders2016a,masters2014}. However, \citet{jones2015} and \citet{sanders2016b} present
evidence that the relationship between [Ne{\sc iii}]/[O{\sc ii}] and directly determined $12+\log(\mbox{O/H})$
does not evolve strongly out to $z\sim 3$. We therefore conclude that the [Ne{\sc iii}]/[O{\sc ii}] indicator should
provide a reliable estimate of the gas-phase metallicity in GOODSN-17940. Accordingly,
our observations provide the {\it first}
nebular oxygen abundance at $z>4$ based on a rest-frame optical spectrum. 

In some respects, the significantly sub-solar metallicity
of GOODSN-17940 is unexpected. In the local universe, there is an observed correlation between dust extinction
and metallicity \citep{heckman1998} in starburst galaxies, suggesting that the dust-to-gas ratio is correlated with the degree
of chemical enrichment. \citet{reddy2010} demonstrated that $z\sim 2$ star-forming galaxies follow
the same correlation between extinction and metallicity, with a similar fraction of metals locked in dust grains.
Accordingly, if GOODSN-17940 followed the same pattern between extinction and metallicity,
its significant dust reddening inferred from both the SED model shown in Figure~\ref{fig:17940-sedhst}
and the H$\alpha$/H$\gamma$ Balmer decrement would suggest 
solar, not $\sim 20$\% solar, metallicities. Investigating the connection between dust extinction and
metallicity at $z>4$ with a larger sample will provide important insights into dust production 
mechanisms in the early universe, and help us place GOODSN-17940 into context.

\section{Discussion}
\label{sec:discussion}

\subsection{Is GOODSN-17940 an AGN?}
\label{sec:discussion-AGN}
The linewidth measured from [Ne{\sc iii}], the highest S/N feature in the GOODSN-17940
spectrum, is $\sigma_v=160$~\kms.  The width of this feature
is larger than those reported in studies of $z\sim3$ ionized gas kinematics
\citep[e.g.,][]{turner2017}. 
Ionized gas associated with an AGN could cause a broad line profile.
If the emission lines in the spectrum of GOODSN-17940 originate from gas excited
by the ionizing spectrum of an AGN rather than massive stars, our estimates
of the nebular metallicity and SFR of GOODSN-17940 are invalid.

However, based on the current evidence, GOODSN-17940 is unlikely to
host an AGN.  The intense star formation in GOODSN-17940 suggests that it is a gas-rich system \citep{schmidt1959}.
If the gas mass fraction in GOODSN-17940 is $\geq 60$\% \citep[easily plausible; see, e.g.,][]{tacconi2013},
applying the virial theorem to the total baryonic mass and size of $\sim 1 \; \rm{ kpc}$ yields
a velocity dispersion consistent with the observed value. Furthermore, the extreme SFR surface density in GOODSN-17940
places it well above the threshold
for driving significant gas outflows \citep{heckman2002}, which could broaden the nebular line profiles \citep[e.g.,][]{genzel2011}.
Finally, we find that GOODSN-17940 is undetected in the 2 Ms {\it Chandra} image of the GOODS-N field,
with a rest-frame $2-10$~keV upper limit of $\log (L_X/\mbox{erg s}^{-1}) = 42.68$ (95\% confidence). 
In the future, measurements of the H$\beta$, [O{\sc iii}]$\lambda 5007$, H$\alpha$, and [N{\sc ii}]$\lambda6584$
features with {\it JWST} will provide stronger constraints on possible AGN contribution.

\subsection{Evolution in the Mass-Metallicity Relation}
\label{sec:discussion-mzr}
Over a wide range of redshifts, star-forming galaxies follow an evolving correlation
between gas-phase oxygen abundance and stellar mass -- the so-called mass-metallicity
relation (MZR) \citep[e.g.,][]{tremonti2004,zahid2014,sanders2015,onodera2016}. At a given
stellar mass, the MZR evolves towards lower metallicity as redshift increases.
With the measurement of [Ne{\sc iii}]/[O{\sc ii}], we can place GOODSN-17940 in the context
of the evolving MZR. For this analysis, we compare GOODSN-17940 with samples
of star-forming galaxies at $z\sim 0, 2, \mbox { and } 3$, both in the empirical
sense of [Ne{\sc iii}]/[O{\sc ii}] vs. $M_*$ and the more physical dimensions of $12+\log(\mbox{O/H})$
vs. $M_*$. For the local comparison sample, we use the $z\sim 0.08$ SDSS composite spectra of \citet{andrews2013},
which are binned by stellar mass. At higher redshift, we draw galaxies from the MOSDEF
survey, again using composite spectra in bins of stellar mass. The two MOSDEF samples
have median redshifts of $z=2.29$ and $z=3.23$. The [Ne{\sc iii}]/[O{\sc ii}] ratio and error are measured for
each composite spectrum, as well as the median stellar mass in the bin. 
In Figure~\ref{fig:ne3o2-m-panels}
(left), a clear progression is evident at fixed mass towards higher [Ne{\sc iii}]/[O{\sc ii}], as redshift
increases from $z\sim 0$ to $z>4$. Adopting the \citet{maiolino2008} calibration between
[Ne{\sc iii}]/[O{\sc ii}] and oxygen abundance (Figure~\ref{fig:ne3o2-m-panels}, right), 
we see that the increase in [Ne{\sc iii}]/[O{\sc ii}] from $z=0-4$ at the mass of GOODSN-17940 
corresponds to a decrease in $12+\log(\mbox{O/H})$ of 1.0~dex. 

In the local universe, the MZR is observed to  have a significant secondary dependence
on SFR, such that, at fixed stellar mass, galaxies are offset towards lower metallicity
at increasing SFRs. The relationship among metallicity, stellar mass, and SFR has been
termed the ``fundamental metallicity relation" (FMR) \citep{mannucci2010}, and has been
interpreted using a scenario where infalling metal-poor gas both stimulates star formation
and dilutes the metallicity. Results thus far are inconclusive regarding the existence of
the FMR at high redshift \citep[see, e.g.,][]{sanders2015,belli2013}. However, if the FMR
does persist out to $z\sim 4$, GOODSN-17940 would be offset towards even lower metallicity
than typical $z\sim 4$ star-forming galaxies of the same mass, given that its SFR is elevated
by an order of magnitude relative to the $z\sim 4$ star-forming main sequence. Accordingly, the
evolution in metallicity out to $z\sim 4$ may not be as extreme as GOODSN-17940 suggests on
its own. It will be possible to resolve this question based on statistical samples of $z\sim 4$ star-forming
galaxies with stellar mass and metallicity measurements from {\it JWST}.

\subsection{Summary and Outlook}
\label{sec:discussion-outlook}
We have presented the first measurements of multiple rest-frame optical
emission lines in the spectrum of a galaxy at $z>4$. Observed 
with Keck/MOSFIRE during the MOSDEF survey, this galaxy, GOODSN-17940, exhibits remarkable properties.
Its active star formation places it significantly above the $z=4$ SFR-$M_*$
main sequence, and, although showing evidence for considerable dust extinction,
its nebular oxygen abundance is only 20\% solar. 

The thermal infrared background prevents
us from obtaining longer-wavelength rest-frame optical spectra from the ground
for GOODSN-17940. However, with the launch of {\it JWST}, it will become commonplace
to analyze the rest-frame optical spectra of $z>4$ galaxies. GOODSN-17940
will make for a particularly compelling target, as it should exhibit detectable
auroral [O{\sc iii}]$\lambda 4363$ emission. Such measurements enable a determination of the electron
temperature, the so-called ``direct" oxygen abundance, and therefore an empirical calibration
between strong emission-line ratios and metallicity. We infer the detectability of [O{\sc iii}]$\lambda 4363$
by calculating the ratio of [O{\sc iii}]$\lambda 4363$/H$\gamma$ emission flux in the 
composite spectrum from \citet{andrews2013} that shows the same [Ne{\sc iii}]/[O{\sc ii}]
ratio as GOODSN-17940. Assuming the same [O{\sc iii}]$\lambda 4363$/H$\gamma$ ratio
for GOODSN-17940, we predict F([O{\sc iii}]$\lambda 4363$)=$4\times 10^{-18}\mbox{ ergs s}^{-1}\mbox{ cm}^{-2}$,
which will be easily detectable with the {\it JWST}/NIRSpec instrument. Finally, we note
that the spectroscopic measurements presented here will be possible out to $z\sim 10$
using {\it JWST}/NIRSpec, well into the reionization epoch.

\section*{Acknowledgements}
We acknowledge support from NSF AAG grants AST-1312780, 1312547, 1312764, and 1313171, grant AR-13907
from the Space Telescope Science Institute, and grant NNX16AF54G from the NASA ADAP program. 
JA acknowledges support from ERC Advanced Grant FEEDBACK 340442.
We also acknowledge the 3D-HST collaboration, who provided us with spectroscopic
and photometric catalogs used to select MOSDEF targets and derive
stellar population parameters.  We thank Daniel Stark for insightful comments.
We wish to extend special thanks to those of Hawaiian ancestry on
whose sacred mountain we are privileged to be guests.


\end{document}